# POLICY–PRACTICE CONTRADICTION: CASE OF CLOUD COMPUTING ADOPTION IN THE MALAWI HEALTH SECTOR

Deborah Amos Phiri, University of Malawi, deborahphiri4@gmail.com

Chipo Kanjo, University of Malawi, chipo.kanjo@gmail.com

**Abstract:** This paper examines the dynamics of policy implementation and how policy contradicts reality on the ground when it comes to practice. The paper finds that despite having well-laid out policy; the actual practice is contrary. Taking data storage policy within the Ministry of Health in Malawi as a case study, the paper highlights that the contextual realities of where Ministry of Health (MoH) data is stored depends on a number of Technology-Organizational-Environmental (TOE) factors. In the wake of cloud computing; some of these factors act as causative factors for data to be stored in the cloud; contradicting the data storage policy.

## 1. INTRODUCTION

A policy provides an overarching framework that is to be enforced by practice. Anderson (2003) defines policy as "a relatively stable, purposive course of action followed by an actor or set of actors in dealing with a problem or matter of concern" (Anderson, 2003; pp.2). This definition distinguishes a policy from a decision, which is essentially a specific choice among alternatives; instead, it interprets policy as something that unfolds over time. The world is full of policies, and they occur at various levels of interaction; personal, organizational, and public. Practice on the other hand is repetition of an activity to improve skill. Policy needs to be translated into activities that constitute practice for it to be effective. However, it is not always the case that what has been articulated in a policy is put into practice. The challenges of policy implementation and success lie in the heterogeneity of the players (Walsham, 2010); as different players may have different agenda. In this paper, we examine the Ministry of Health (MoH) data storage policy which stipulates that all patient-related data should be stored within the borders of Malawi.

The policy comes at a time when most organizations are turning to cloud computing for storage. Cloud computing is a model for enabling universal, convenient, on-demand network access to a shared pool of configurable computing resources such as networks, servers, storage, applications, and services (Mell & Grance, 2011). In the recent decade Malawi health sector has shifted from traditional ways of manual collection, storing and managing health data to digital technologies due to the growing demand for good quality health information and also as a result of performance-based resource allocation by donors. This shift has brought about the need to come up with effective ways of storing, sharing and access to the data that should also keep up with the growing demands of the data. Evidence has shown that cloud computing use in the healthcare offers many benefits such as storage of large amounts of data (it is scalable and elastic for increasing or decreasing), offering remote data access and allowing sharing between authorized units (Manya, Nielson, & Pundo, 2016).

The Malawi National Health Information Systems Policy (2015) states that any health-related data whether physical or electronic shall be stored only within the borders of Malawi except for the purpose of continuation of care, on the other hand there is evidence of cloud computing utilization in the health sector. This paper describes and analyses the policy-practice contradictions when it comes to implementation. Contradiction here is defined as the dynamic interplay between unified opposites that are actively incompatible (Baxter & Montgomery, 1998) and contrary to each other.

Malawi health sector was chosen as the context of study because the resources necessary for information management in the health sector such as skilled human resource, finance, technical and physical infrastructure (Smith, 2018), laws and policies have not evolved in parallel to cope with increased demand for data management (MoH, Malawi National Health Information System Policy, 2015) a condition that may force the country to adopt cloud computing technology.

## 2      LITERATURE REVIEW

### 1.1    Why Policy

A Policy is important because no public activity can be attempted without the stipulation of clear objective and a proper policy (Marume, 2016). As such, public policies emerge in response to policy demands, or those claims for action or inaction on some public issue made by other actors (Anderson, 2003). Other than the definition of policy provided in the introduction section above, Dodd and Boyd (2000) had defined policy as a plan of action agreed to by a group of people with the power to carry it out and enforce it. In their definition, two terms are of interest: having the power to carry it and enforcing the policy. Policy plays an important role in influencing the degree which research findings may influence health services, however, while there may be extensive research on the effectiveness of health-care interventions, there is often less evidence on their cost-effectiveness, implementation, cultural appropriateness and effects on health inequalities, all of which are important considerations for policy-making ().

### 1.2    Factors affecting Policy Implementation

A number of factors affect policy implementation, Kunyenje (2019) found that excessive influences of the external actors led to the resulting National ICT policy not being owned and consequently not implemented. Social challenges such as shortage of infrastructure (MoH, Malawi National Health Information System Policy, 2015) also affect policy implementation.

### 1.3    Cloud Computing in the Health Sector

Cloud computing is increasingly becoming important in the generation, storage and transmission of information worldwide and has made a tremendous impact on the information technology industry over the past years (Manya, Nielson, & Pundo, 2016). The health sector is one of the fields that is gradually gaining acceptance of cloud computing technology as an effective means of improving healthcare delivery across the globe (Mgozi & Weeks, 2015). The effectiveness of health care services relies on the availability of information systems that convey timely, accurate and readily available information to health practitioners, policy makers, researchers and the general public (Yi et al., 2008) as cited in (Msiska, 2018).

Institutions do not need to invest on hardware, complex technical capacity and maintenance because these concerns are already taken care of by the cloud computing providers thus resulting in a more efficient and improved health care management (Munene & Macharia , 2015). Cloud computing essentially stores all of its applications and databases in the data centers which are mostly stationed in different locations. Initiatives in eHealth should be seen as an investment in the health sector to secure benefits that exceed costs over time by adopting appropriate architecture coupled with comprehensive and rigorous interoperability standards to ensure sustainable HIS (Mgozi & Weeks, 2015). It should also be known that despite the cloud computing benefits there are also some concerns mostly to do with data ownership, privacy and security of the data due to the location of the servers by many institutions (Rieger, Gewald, & Schumacher, 2013; Sahay, Sundararaman, & Braa, 2017).

### 1.3.1 Key Benefits of Cloud Computing

There are different ways in which cloud computing may benefit the health sector and make health industry more advanced and efficient. These include: *pay-as-you-grow* where public cloud providers like Amazon allow companies to avoid large up-front infrastructure investment. This model is a relevant approach in health institutions in developing countries where financial resources for investment in hardware are scarce and the human capacity to maintain local servers is often lacking and they cannot afford to spend large sums of money at the beginning of their business journey (Manya, Nielson, & Pundo, 2016). A second benefit is *initial cost saving*, this eliminates upfront cost, offering economic savings and financial benefits of decreasing the cost of entry by reducing the initial and operational costs for start-up of new IT projects (Abubakar, Bass, & Allison, 2014; Alharbi, Atkins, & Stanier, 2017). *Simplicity* is another key benefit where there is relief of not having to buy and configure new equipment. This allows institutions and their IT staff to get right to their core business as the cloud solution makes it possible to get the application started immediately, and it costs a fraction of what it would cost to implement an on-site solution (Velte, Velte, & Elsenpeter, 2010). Finally, there is *accessibility of information*; Cloud computing enables remote access to information and applications irrespective of the location and accessible via the internet using a secure authentication (Council, 2017). Literature shows that cloud in health opens up a new horizon facilitating availability of information irrespective of the location of the patient and the clinicians which enables improved clinical outcomes and physicians make right decisions when critical information such as blood types, X-Ray and test results are viewed at the right time (Priyanga.P, 2015; Almubarak, 2017).

### 1.3.2 Challenges of Cloud Computing

As with any other technology cloud computing has its own challenges that play a major role in the decision-making process for institutions. It should be well known that challenges of cloud computing differ between different context. *Data security* being the most challenging barrier to cloud computing in that some institutions fear its adoption due to the sensitivity of their data. Security is linked with the concepts of integrity, confidentiality, authenticity and availability which should be carefully examined (Anand, 2018). This is the most dominant factor influencing the adoption of cloud computing particularly in health institutions because the data requires a more secure environment for storage and retrieval of information (Lian, Yen, & Wang, 2014).

However, there are two schools of thought, one as articulated above, argue that given the sensitive nature of the health information, there may be some reservations because everyone wants to be assured that their private and confidential data is safe. A second school of thought argue that there is no guarantee that data is better protected internally compared to the cloud; in a sense that there is a possibility that data could be even safer in the cloud because cloud providers may pose higher level of data security expertise compared to in-house expertise (Kuo, 2011; Gorelik, 2013; Abubakar, Bass, & Allison, 2014). *Lack of awareness* of what cloud computing concepts are and what benefits and risks exists when institutions adopt is another challenge especially in developing countries, and this may weaken its adoption (Muhammed, Zaharaddeen, Rumana, & Turaki, 2015). The lack of exposure and prior experience makes it impossible for top management IT personnel to be fully aware of what it entails to adopt cloud computing technology (Dahiru, Bass, & Allison, 2014; Abubakar, Bass, & Allison, 2014; El-Gazzar, Hustad, & Olsen, 2016). Another challenge is *vendor lock-in*, a situation where customers are dependent (i.e. locked-in) on a single cloud provider technology implementation and cannot easily move to a different vendor without substantial costs, legal constraints, or technical incompatibilities (Opara-Martins, Sahandi , & Feng , 2016). Applications that are developed on a platform of one cloud provider cannot be easily migrated to the other provider as they may be running different versions of the same open-source system and may have their own ways of doing things. Each cloud provider for SaaS solutions creates its own application programming interfaces (APIs) to the application (Petcu & Crăciun, 2011). Since there are no uniform cloud management software standards across different providers available, once an institution invests time and resources to establish operations on one cloud platform, it can be difficult to switch (Gorelik, 2013).

### 1.4 Applications of Cloud Computing in Health

Literature indicates that some countries are already using cloud computing services in health. A South African non-governmental organization, combines the cloud with database technology and mobile services to fight HIV/AIDS transmission from mother to children. They digitize patient records and shares them with counsellors across its networks of over 700 sites in Africa. The records contain information on treatment plans, and advanced reporting tools, which allow quick response (Kshetri, 2011). In Kenya, cloud computing is used to host a web-based national data warehouse. This uses an international cloud hosting company physically located in London for implementation of the DHIS2 system. The study showed that with that development the immediate benefit was availability of data, which made it possible for stakeholders at all levels of the government to access the data online avoiding bureaucracy of seeking permission from the decentralized government units. This, in the long run also enabled the integration of national data thus reducing fragmentation of parallel systems. (Manya, Nielson, & Pundo, 2016). Furthermore, a study showed that in Australia, cloud computing is used in telemedicine, combined with multiple concepts such as e-appointment, e- consulting, and e-prescription to enable patients to use the internet for maintaining remote connection with their physicians and discussing their health-related problems (Moghaddasi & Tabrizi, 2017). The resulting system allowed the physician to easily access the patient's medical history, files, test results, etc (*ibid.*).

### 1.5 Factors Influencing Cloud Computing Adoption

There are several factors that may influence institutions to adopt cloud computing despite the known characteristics and benefits that the technology offers. Cloud computing is not a one size fits all hence institutions need to assess their current and future IT systems needs very well and understand its concepts before making any decision to adopt the technology or not. Cloud computing adoption in the health institutions requires strategic planning and insight to gain full advantages and trade-offs of this new model (El-Gazzar, Hustad, & Olsen, 2016). These factors have been grouped in terms of technological, organizational, and environmental as stipulated in the TOE theoretical model.

#### 1.5.1 Technological Factors

The technological context includes all of the technologies that are relevant to the firm, both technologies that are already in use at the firm as well as those that are available in the marketplace but not currently in use (Baker, 2011). Technological factors comprise of relative advantage, complexity, and compatibility.

### 1.5.2 Organizational Factors

The organizational context includes attributes such as size, quality of human resources, and complexity of the firm's managerial structure (Low, Chen, & Wu, 2011). Top management support and behaviors are critical for creating a conducive environment for providing adequate resources for the adoption of new technologies as cited in (Low, Chen, & Wu, 2011). Top management must be willing to allocate valuable organizational resources (Willcocks & Sykes, 2000). In addition, top management must have sufficient information for cloud services which it requires and align with the vision of their institution if they are to use a specific cloud service (Abdollahzadehgan, Hussin, Gohary, & Amini, 2013).

### 1.5.3 Environmental Factors

Environmental factors refer to a firm's industry, competitors and government policy or intention (Low, Chen, & Wu, 2011). The environmental factors that have been used in this study are government policy and vendor scarcity. *Government Policy* refers to policy relating to the geographic location of where the data is stored. Data protection is a general concern for most countries wishing to adopt cloud computing technology. a study conducted in South Africa show that the legislative policy framework is a critical enabling factor to resolve key concerns inhibiting the adoption of cloud computing in the health sector (Mgozi & Weeks, 2015). Whereas the German banking sector revealed that government regulations are a major factor influencing the cloud-computing decision as they are highly interdependent with other factors like security and compliance requirements ,the data and provider location (Rieger, Gewald, & Schumacher, 2013). Furthermore, a study conducted in Saudi Arabia revealed that national regulation about the implementation of cloud computing can play an important role in supporting or slowing the adoption (Alharbi, Atkins, & Stanier, 2017). *Vendor Scarcity* is another environmental factor which is the availability of credible and experienced cloud service providers. A study conducted in China revealed that institutions may be constrained in their vendor selection if the service they need is not available (Li, Zhao, & Yu, 2014). In German, a study showed that the image of the cloud service provider plays a role as it builds trust, especially for handling sensitive data (Rieger, Gewald, & Schumacher, 2013) and the cloud provider's reputation plays an important role in meeting business needs as stipulated in the service level agreement (Sadoughi, Ali, & Erfannia, 2019).

### 1.6 Policy Diffusion

*Policy diffusion* is a process where policy innovations spread from one government to another (Shipan and Volden 2008). Policy-makers learn from other governments experiences: particularly where an adopted policy is deemed successful elsewhere. The second mechanism, economic competition, can lead to the diffusion of policies with economic spillovers across jurisdictions. Policy-makers consider the economic effects of adoption (or lack thereof). Particularly where there are positive spillovers, governments are more likely to adopt the policy of others (Shipan and Volden 2008). Another mechanism is imitation/emulation which means 'copying the actions of another in order to look like the other' (Shipan and Volden 2008: 842). The last mechanism is coercion, this is different from the other three which are voluntary. For instance, countries can coerce one another through trade practices or economic sanctions, either directly or through international organizations (Shipan and Volden 2008).

### 1.7 Conceptual framework

This study adopts concepts from two theories: i) the TOE theoretical framework (Baker 2011) to illuminate the factors' that influence cloud computing adoption as detailed in section 2.5. The theory focuses on Technological, Organizational and Environmental issues of which policy is one of the specific environmental factors. ii) As a way of understanding how the data storage policy came into being, *policy diffusion* is used to complement the TOE theoretical framework, focusing on the mechanism of imitation/emulation.

## 2    METHODOLOGY

This paper is part of findings of a qualitative study which was chosen with the aim of gaining an in-depth understanding of cloud computing adoption in the context of health institutions. The research was carried out in different health institutions in Malawi which helped in answering different factors influencing cloud computing adoption, the perceived benefits and potential barriers to the adoption of cloud computing at the level of the institution in the health sector. The factors assisted in examining the policy practice contradiction that exists.

Data was gathered from fourteen (14) health institutions targeting middle and top managers who were purposively selected as respondents with experience and expertise in health information systems management and knowledge of cloud computing technology.

Whilst the broader study was guided by the Technology Organizational and Environment (TOE) theoretical framework, this paper complements TOE theoretical framework with Policy diffusion (imitation) concept to be able to understand the genesis of the data storage policy.

### 2.1    Data Collection

The respondents for the study were from fourteen (14) selected health institutions in Malawi that are currently managing different data systems either for program monitoring, studies or clinical trials. The study used purposive sampling technique to choose key respondents, which is the deliberate choice of an informant due to the qualities the informant possesses (Tongco, 2007). The respondents interviewed were mostly middle or top manager representatives as these were perceived to be the most knowledgeable or experts in the research area and perceived to provide relevant information pertaining to the study. The initial sample size for the study was nineteen (19) health institutions out of 31 health institutions registered to be actively using different IT solutions. Nineteen (19) institutions were purposively selected and invitation for interviews were sent via email and followed up by phone, fourteen (14) responded and accepted to take part in the interview. All the interviews were done virtually due to covid-19 which made it impractical to arrange for face-to-face interviews or follow-up.

### 2.2    Approaches to Theory

This study was an interpretive approach and was guided by a Technology Organizational and Environment (TOE) theoretical framework. The data collected was guided by the main themes that are used in the framework namely *technology* which the researcher looked in terms of relative advantage, complexity and compatibility, *organization* which the researcher looked in terms of top management support, technology readiness particularly IT infrastructure as well as the quality of human resources and lastly *environment* in terms of government policies and vendor scarcity.

Table 1 – Respondents' communication mode

| No. | Institution | Position | Communication Mode | Found Via |
|---|---|---|---|---|
| 01 | R1 | Health Informatics Manager | Skype | LinkedIn |
| 02 | R2 | IT Manager | Skype | LinkedIn |
| 03 | R3 | IT Officer | WhatsApp Call | Other Colleagues |

| 04 | R4  | Country Director | Email | Colleagues |
|----|-----|------------------|-------|------------|
| 05 | R5  | Senior IT Officer | Skype | Colleagues |
| 06 | R6  | Software Development Manager | WhatsApp Call | Colleagues |
| 07 | R7  | Health Information Systems Officer | WhatsApp Call | Colleagues |
| 08 | R8  | Head of Data Unit | Skype | LinkedIn |
| 09 | R9  | Data Core Leader | Skype | Colleagues |
| 10 | R10 | Project Manager | Email | Colleagues |
| 11 | R11 | IT Specialist | Skype | Colleagues |
| 12 | R12 | IT Officer | Email | Other Colleagues |
| 13 | R13 | Senior Technical Advisor | Skype | Other Colleagues |
| 14 | R14 | Consultant | WhatsApp Call | Other Colleagues |

## 2.3 Data Analysis

The first five (5) recorded interviews were manually transcribed and the remaining ones were transcribed using Express Scribe software version 9.09 upon the researcher being introduced to it, which proved to be faster than the manual transcription. Data analysis was done based on the main themes that were used in the theoretical framework.

## 3 FINDINGS AND ANALYSIS

The study found that cloud computing was already in use in most of the institutions interviewed. Eleven (11) institutions out of the fourteen (14) interviewed were already using cloud services in their institutions and one (1) once used but was no longer using and the other two (2) were not using any cloud services. Table two shows the different cloud services used by the institutions interviewed.

*Table 2: Cloud Usage in Institutions*

| Institution | Core Competence | Type of Institution | Using Cloud | Type of Service | Cloud Usage |
|-------------|-----------------|---------------------|-------------|-----------------|-------------|
| R1 | ● Health SystemStrengthening | International | Yes | SaaS, IaaS | ● Data hosting<br>● Online collaboration |
| R2 | ● Primary Health Care | International | Yes | SaaS, PaaS | ● Data hosting<br>● Build applications<br>● Online collaboration |
| R3 | ● Health Systems Strengthening | National | No | N/A | N/A |

| R4 | ● HIS Innovation | International | Yes | PaaS | ● Hosting applications |
|---|---|---|---|---|---|
| R5 | ● Healthcare and systems delivery | International | Yes | SaaS, IaaS | ● Data storage ● Online collaboration |
| | ● | | | | |
| R6 | ● HIS Innovation | National | No | N/A | N/A |
| R7 | ● Healthcare service delivery | National | Yes | PaaS | ● Data collection |
| R8 | ● Clinical Research | International | Yes | SaaS, IaaS | ● Data backup ● Online collaboration |
| R9 | ● Epidemiological Research | International | Yes | PaaS, IaaS | ● Data collection ● Data hosting ● Build applications |
| R10 | ● Capacity building and Research ● HIS Innovation | National | Yes | SaaS, IaaS | ● Data hosting ● Online collaboration |
| R11 | ● Health Systems Innovation | International | Yes | SaaS, PaaS | ● Data management ● Build applications |
| R12 | ● Healthcare service delivery | National | No | N/A | N/A |
| R13 | ● HIS Strengthening | National | Yes | IaaS | ● Data hosting |
| R14 | ● HIS Strengthening ● HIS Innovations | International | Yes | PaaS | ● Hosting applications |

The study shows that some factors influencing cloud computing adoption are context specific to developing countries and a lot of investment must be done as well as create awareness campaigns to enlighten institutions on the risks and benefits associated with the technology to clear out any misconception that institutions may have.

## 3.1   Is the Malawi Health Sector Technological Ready?

Most respondents mentioned that for their IT human resources there is readiness. However, lack of basic IT and network infrastructure was their main hinderance in the adoption of cloud because that is a prerequisite for cloud computing. Poor ICT infrastructure and poor connectivity were also mentioned as technological factors affecting cloud computing adoption.

> *The most common barrier which I have noticed in these health centers is usually connectivity, so most of the health centers are located in the remotest part of this country where you cannot even make a phone call, connectivity is quite a challenge"* **R2.**

### 3.2 Vendor Scarcity

The availability of credible vendors was another factor that this study found to significantly influence the level of adoption of cloud computing. Respondents stated that there is scarcity of cloud vendors within Malawi, a thing that leads them to use other cloud vendors outside Malawi. Some respondents mentioned that having a reliable cloud provider especially within the country would increase trust and enable all those with doubts about the technology to physically go and ask. At the time of write-up only one cloud provider was identified in Malawi. In addition, because the credibility and history of the new entrants in the cloud service market is not yet known, it may have a negative impact on the level of confidence that institutions may have.

> *"We will wait and see what others will experience with this new vendor, we cannot give them all our data or give them business when we do not know their history, we do not want to be part of the experiment, we would rather be with those that are already known"* **R5.**

### 3.3 Government Policy

Government policies related to IT in this study showed that it is a significant factor that may inhibit the adoption of cloud computing. The location of where the data is replicated for disaster recovery preparedness is something that may break the regulations put forward by the Malawi government on the geographical boundary of the data. The results in this study are consistent with the findings from studies conducted in the German banks and Bangladesh education sector that showed that government regulations heavily influence the cloud computing decision in that the location where data is being stored and where the cloud provider resides is a major factor influencing cloud computing decision-making (Rieger, Gewald, & Schumacher, 2013; Rahman & Rahman, 2019). Furthermore, the respondents were asked on their perception on the current government policies and if they are at par with current HIS innovations. The respondents had mixed views; others expressed that the policies are at par with current HIS innovations but only lack enforcement but 36% of the respondents felt that they need to be updated because ICT has evolved. A similar study conducted in Malawi on cloud computing adoption in Higher Education Institutions (HEI) also concurs with the findings of this study that policies and regulations put forward by the government may affect the environment for the adoption of cloud computing in institutions (Makoza, 2016). Potential Benefits for Cloud Computing Adoption

The respondents (50%) expressed that cloud computing can offer cost saving in institutions by transitioning from capital expenditure to operational expenditure in form of monthly or annual fees. Storage capacity that comes with the cloud technologies was another benefit cited by 37% of the respondents. This was viewed from the volumes and variety of data that most institutions are storing for different programmes. One (1) respondent expressed the storage capacity to be one of the gains for cloud technologies in this regard: *"The potential benefits that we can have is the storage capacity because I know at the […] we have limited capacity of hosting large data, so we can gain on that"* **R3.** It was also mentioned that cloud can act as a means of backing up data in case of unforeseen circumstances such as loss of data by the institutions.

Broad network access where data or applications are accessed anytime from anywhere from any device with internet access was another benefit that respondents mentioned. Institutions that have offices outside of Malawi or those within Malawi but operating in different regions or districts shared this benefit. Another benefit cited was system availability. System availability for cloud infrastructure can be referred to the time that the data center is accessible or delivers the intended IT service.

### 3.4  Potential Barriers of Cloud Computing Adoption

The barriers to cloud computing have been summarized as indicated in Table 2.

Table 3 - Barriers to Cloud computing

| Barriers of Cloud Computing | Responses |
|---|---|
| Connectivity | Eight (8) respondents expressed concern on how poor the connectivity is for Malawi to support a technology like cloud computing. |
| Data Security | Six (6) respondents highlighted that the nature of data that most institutions have and the fear of privacy and confidentiality breach when they move it to a cloud storage is a barrier. |
| Running Costs | Six (6) respondents expressed a concern on the running cost of cloud such as high internet charges to be the barrier for cloud adoption |
| Lack of Awareness | Four (4) respondents mentioned that lack of awareness of cloud computing weakens the adoption. |
| Perception | Three (3) respondents mentioned that the perception that people have that they will lose control of their data and that it is not safe if moved to cloud is more likely to hinder its adoption |
| Government Policies | Two (2) respondents mentioned that the government policies are not accommodating to support cloud technologies |
| Legislations | One (1) respondent expressed that the legal differences in place by different regions hosting cloud technologies may be a barrier for cloud computing |

## 4   DISCUSSION AND CONCLUSION

### 4.1  Technological Readiness

Most respondents mentioned that for their IT human resources there is readiness. However, lack of basic IT and network infrastructure was their main hinderance in the adoption of cloud. This is consistent with findings from study conducted in Portugal on the determinants of cloud computing adoption (Oliveira, Thomas, & Espadanal, 2014) which stated that technology readiness is a significant factor and that institutions must ensure that the technology infrastructure as well as IT specialists are adequate for the integration of cloud-based solutions in their institutions. However, this study also highlighted some context-specific factors such as poor ICT infrastructure and poor connectivity which emphasize findings in most studies related to developing countries (Muhammed, Zaharaddeen, Rumana, & Turaki, 2015; Skafi, Yunis, & Zekri, 2020).

### 4.2  Vendor Scarcity

The availability of credible vendors was found to significantly influence the level of adoption of cloud computing. Respondents stated that the scarcity of cloud vendors especially within Malawi may lead institutions to engage external vendors thus resulting into non-compliance with what the policy stipulates. A study conducted in China (Li, Zhao, & Yu, 2014) also found that institutions may be constrained in their vendor selection if the service they need is not available with existing vendor because of lack of competition in the emerging market, which may compromise the quality of services provided. A study conducted in a Jordanian Hospital articulates that vendor scarcity plays a vital role, given that the presence of sufficient, reputable, and competent vendors will encourage institutions to adopt cloud services and create trust with the providers and the technology itself (Harfoushi, Akhorshaideh, Aqqad, & Janini, 2016). Here we have a situation where we have only one cloud provider as indicated in the findings whose history or reputation is not yet established which may have an impact on the choice of provider by clients. The global cloud service providers have more and secure data centers and are more careful when providing cloud computing services making sure they do it in an appropriate manner (El-Gazzar R. F., 2014).

### 4.3   Policy, Practice and Pragmatism

The Malawi government has a policy that stipulates that any patient-related data generated locally must be housed within the geographical boundaries of Malawi. The findings of this study, however, show that seventy six percent (76%) of the institutions interviewed were already using the cloud services. One (1) of the respondents further stated that they use cloud storage because it was a prerequisite put forward by the donor funding their project but further expressed that the data they store for such projects is not sensitive. Two (2) respondents mentioned that they use cloud computing for easy data sharing and collaboration with other staff outside Malawi. In addition, from secondary data collection it was noted that 70% of implementing partner institutions host their data off-premise outside Malawi. This is a clear indication that what the policy stipulates and reality on the ground is contrary, at the time of data collection there was no indication of any a single enforcing policy decision into implementation concerning cross-border data transfers. Dodd and Boyd (2000) had defined policy mentioning two terms of interest in their definition: i) having the power to carry it and ii) enforcing the policy. Where donor puts forward conditions where to store data, then the power to carry out is lost. Additionally, the findings clearly indicate that over 70% are already using the cloud, a thing that makes enforcement difficult to implement.

The practicality of having policy enforcement match practice on the ground would be to have a central data repository locally being managed by the available local cloud provider. In addition, there must be a supporting updated national policy to establish a quality and secure cloud hosting environment in these local data centers. Having a central repository would also ensure that data remains in the country even when some international projects phase out.

However, with the performance-based funding allocation put forward by most donors, cloud technology may be one way that institutions looking for solutions that will enable them to monitor their data in a timely manner from anywhere.

### REFERENCES AND CITATIONS